\documentclass[runningheads]{llncs}

\usepackage{amsmath}

\usepackage{graphicx}
\usepackage{float}
\usepackage[usenames,dvipsnames]{xcolor}

\newcommand{\sumi}{\sum_{i=1}^N}

\newcommand{\ham}{\mathcal{H}}
\newcommand{\lag}{\mathcal{L}}
\newcommand{\qq}{{\bf q }}
\newcommand{\ppp}{{\bf p }}

\definecolor{HSafron}{RGB}{232,125,30}

\begin{document}

\title{Physical Symmetries Embedded in Neural Networks}
\author{M.  Mattheakis\inst{1},  P. Protopapas\inst{1}, D. Sondak\inst{1}, M. Di Giovanni\inst{2},   E. Kaxiras\inst{1,~ 3}}

\authorrunning{M. Mattheakis et al.}
%
\institute{SEAS, Harvard University, Cambridge, MA, United States   \and Polytechnic of Milan, Italy \and Department of Physics, Harvard University, Cambridge, MA, United States\\
\email{mariosmat@g.harvard.edu}
}

\maketitle

\begin{abstract}
Neural networks are a central technique in machine learning.  Recent years have seen a wave of interest in applying neural networks to physical systems for which the governing dynamics are known and expressed through differential equations.  Two fundamental challenges facing the development of neural networks in physics applications is their lack of interpretability and their physics-agnostic design.  The focus of the present work is to embed physical constraints into the structure of the neural network to address the second fundamental challenge.  By constraining tunable parameters (such as weights and biases) and adding special layers to the network, the desired constraints are guaranteed to be satisfied without the need for explicit regularization terms.  This is demonstrated on supervised and unsupervised networks for two basic symmetries:  even/odd symmetry of a function and energy conservation.  In the supervised case, the network with embedded constraints is shown to perform well on regression problems while simultaneously obeying the desired constraints whereas a traditional network fits the data but violates the underlying constraints.  Finally, a new unsupervised neural network is proposed that guarantees energy conservation through an embedded symplectic structure.  The symplectic neural network is used to solve a system of energy-conserving differential equations and out-performs an unsupervised, non-symplectic neural network. 
\end{abstract}

\section{Introduction}
Machine learning (ML) has become very popular due to its significant contributions to a variety of scientific fields including the natural and social sciences, medicine, and finance.  Although ML consists of many algorithms, neural networks (NN) have emerged as the de facto standard for some tasks including image recognition and natural language processing.  Because of their success in these fields, there is great interest in using NNs to make predictions in science and engineering.
 Supervised NNs have been used to improve turbulence models \cite{julia2016a,julia2016b,rui2019}, predict material properties \cite{prl2018,trevor2019}, solve quantum many body problems \cite{manybody2017}, forecast the future behavior of dynamical systems \cite{Ott2017,sapsis2018,marios2018}, solve differential equations (DEs) \cite{karniadakis2017jocp,sinai2018}, and even discover DEs \cite{kardiadakis2017jocp_b,kutz2017data,perdikaris2019}.  New network architectures for solving DEs are an active area of research. 
  The neural ordinary differential equations network~\cite{nips2018rc} is a particularly interesting interpretation of NNs.  Recasting a NN as a continuous ordinary differential equation (ODE) was shown to result in large performance gains in computational time and memory footprint.  Unsupervised NNs have also been used to solve both ordinary and partial DEs ~\cite{lagaris1998,spiliopoulos2018,nips2018}.  Because of the universal function approximation property of NNs~\cite{hornik1991}, they may be a natural approach to solving complicated physical problems governed by differential equations.

 Although ML  algorithms, and specifically NNs, have experienced a flurry of development in recent years, there are still several outstanding issues when applying them to problems in the physical sciences.  Two primary concerns drive skepticism within the scientific community of the applicability of NNs  to physical problems.  First, NNs  are  difficult to interpret.  Whereas many classical algorithms, such as the singular value decomposition and Fourier series, are based on hierarchical, orthogonal bases, NNs  lack any such obvious interpretability. They are therefore often treated as a black box technology with no recourse for unpacking how or what they learned.  Interpretable models are extremely useful in physics because they allow scientists and engineers to gain insight to physical processes at different levels of importance. Second, NNs, in their native form, are physics-agnostic.  That is, a NN  may fit a given data set, but it may violate the underlying physical laws that generated that data set in the first place.  Any predictions with such a NN  cannot be trusted in parameter regimes for which there is no data even though the governing physical laws may be the same.  In the present work, we seek to address the second concern by embedding physical constraints directly into the network architecture.  Physical constraints, such as conservation laws, have typically been included as regularization terms during the training process~\cite{perdikaris2019}.  This leads to convenient implementations, but ultimately does not exactly preserve the desired constraints.  Recent approaches for imposing constraints in NNs  include lattice translation~\cite{manybody2017} and embedded Gallilean invariance \cite{julia2016a,julia2016b}. Furthermore, it has been shown that NN structures with embedded constraints inspired by symplectic numerical integrators can remedy issues like the exploding and vanishing gradients in recurrent NNs~\cite{symplecticRC2018}.
 
 The approach taken in the current work directly embeds physical constraints within a NN.  The main idea is the introduction of hub neurons (or a hub layer) that enforce the desired constraint.  The weights and biases of the hub neurons are derived in such a way that predictions from the  NN  are guaranteed to preserve constraints such as even/odd symmetry and energy conservation. In section~\ref{sec:symmetry} we introduce the basic idea of the hub neurons and apply it to regression problems involving functions with even / odd symmetry where the underlying function has been obscured by noise that may destroy the true symmetry.  A hub layer is derived that guarantees that the even / odd symmetry of the underlying system is respected.  Similarly, in section~\ref{sec:energy_regression} the goal is to fit a regression line to noisy data that was generated from an energy-conserving process.  Again, the concept of the hub layer is used to enforce the energy conservation in the final prediction.  Section~\ref{sec:symplectic} focuses on unsupervised NNs  for solving ODEs  that conserve energy.  We introduce a symplectic structure into the network architecture itself, which guarantees that the unsupervised network conserves energy when solving the ODEs.  The paper concludes in~\ref{sec:conclusion} with a summary of the key ideas introduced in this work and a discussion of future plans.


\section{Even/Odd Symmetry}
\label{sec:symmetry}
A prototypical example of symmetry is the even/odd symmetry of a particular function.  An even function $f_{\textrm{even}}\left(t\right)$ has the property that $f_{\textrm{even}}\left(t\right) = f_{\textrm{even}}\left(-t\right)$ while an odd function $f_{\textrm{odd}}\left(t\right)$ has the property that $f_{\textrm{odd}}\left(t\right) = -f_{\textrm{odd}}\left(-t\right)$.  In this section, we consider the situation in which data is generated from a function that is \textit{known} to be even (or odd).  The challenge is that noise in the data may break the known symmetry.  A standard feed-forward multilayer perceptron (MLP) is unaware of the original symmetry and will therefore perform a regression on data that is not symmetric thereby resulting in predictions that violate the desired symmetry.  To counter this issue, we design a hidden layer which we call the \textit{hub} layer. When the hub layer is used as the last hidden layer of an MLP, the regression is guaranteed to be even or odd function as desired.

We consider an MLP that consists of $L$ hidden layers with $N$ neurons per layer and a single linear output node. The network takes one input  $t$ and returns one output $\hat {x}(t)$.  The activation function of a neuron of the last hidden layer is denoted by $h(t)$, which is a composition of all the previous layers 
The output has the form:
\begin{align}
\label{eq:y}
\hat x(t) &= \sumi     w_i ~ h_i(t) + b,
\end{align}
where $b$ is the bias of the output node, $w_{i}$ are the weights of the last hidden layer, and $i$ denotes the index of the neuron in the last hidden layer.  
We guarantee that  $\hat x(t)$ is an odd or even function by demanding that Eq. (\ref{eq:y}) satisfies the relationship $\hat x(t) = s \hat x(-t),$ where $s=\pm 1$ accounts for even and odd symmetry, respectively.  
 Demanding that $x\left(t\right) = -x\left(-t\right)$ in Eq.~\eqref{eq:y} imposes a constraint on the bias of the last node of the form,
\begin{align}
  b = -\dfrac{1}{2}\sumi w_{i} \left[ h_{i}(t) + h_{i}(-t) \right] \nonumber 
\end{align}

\noindent To simplify notation, we denote  $h_i^{\pm} = h_i(\pm t)$ and $H_i^{\pm} = h_{i}^{+} \pm h_{i}^{-}$ and enforcing this relationship on the bias term of   Eq. (\ref{eq:y}) leads to 

\begin{align}
  \hat x(t) & = \frac{1}{2}\sumi w_i H_i^-.
  \label{eq:symodd}
\end{align}
Note that any predictions from a NN  with the embedded constraint on the bias expressed through Eq.~\eqref{eq:symodd} will automatically guarantee odd symmetry in the prediction.  We further note that Eq.~\eqref{eq:symodd} is the odd part of Eq.~\eqref{eq:y}.  More generally, the even-odd decomposition of Eq.~\eqref{eq:y} is,
\begin{align}
    \hat x(t) & = \frac{1}{2}\left( \sumi w_i H_i^+ +2b \right)+\frac{1}{2}\left(\sumi w_i H_i^-\right).
    \label{eq:symForward}
\end{align}
Hence, Eq.~\eqref{eq:symForward} provides a concise ansatz for embedding even/odd symmetry and also motivates the idea of hub neurons, a special nodes that have a special operation after the activation $H_i^{\pm}$.
 The  advantage of networks with the hub neurons  is that when physics requires the observation to be odd or even but the symmetry is broken due to  noise,  we are able to retain the correct symmetry and make predictions that are  physically acceptable. In addition, the hubs reduce the number of the  NN solutions  and thus, the training becomes more efficient.

\begin{figure}[ht]
    \centering
     \includegraphics[scale=0.24]{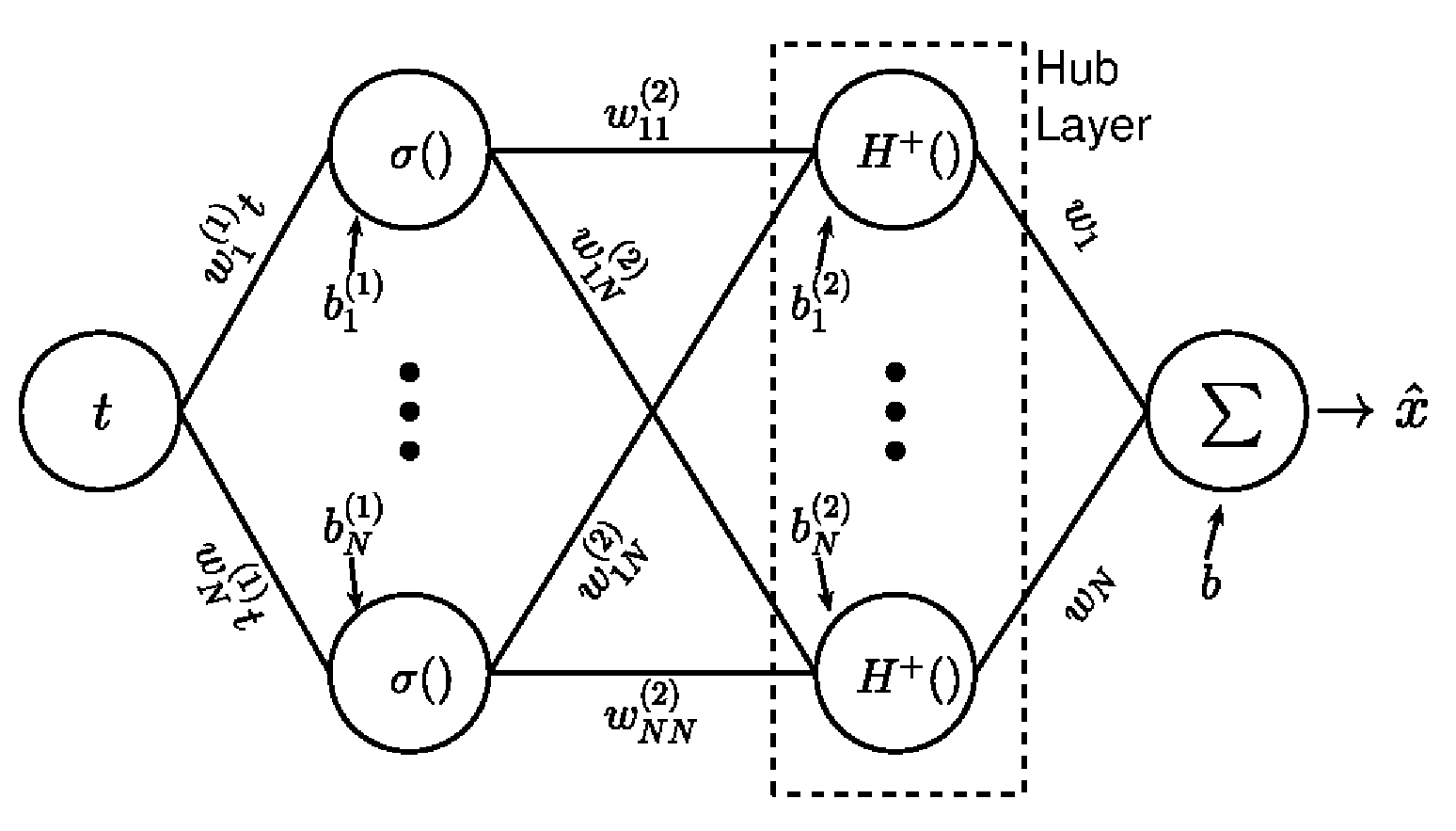}
    \caption{The new NN architecture with an even hub layer to guarantee the underlying even symmetry.    } 
    \label{fig:my_label}
    \label{fig:hubNN}
\end{figure}

To test our method, we generate artificial data from the cosine function (an even function) with Gaussian noise, $\epsilon \sim  {\cal{N}} \left(0, \sigma\right)$,
The data are therefore generated from,
\begin{align}
    \label{eq:cosSrc}
    x(t) = \cos(t ) +\epsilon, \quad t\in\left[-2\pi, \ 2\pi \right] .
\end{align}
 A NN should learn an underlying even function, 
 whereas we expect that a standard MLP will learn a function that does not necessarily preserve the even symmetry.  Figure~\ref{fig:hubNN} depicts the new NN architecture used on this problem with a hub layer, which is given by the first term of the right-hand of Eq. (\ref{eq:symForward}), to preserve the even symmetry.  In subsection~\ref{sec:no-phase} we compare predictions and performance of a standard MLP to the hub network using data generated from an even function for a range of noise levels, $\sigma$.  Subsection~\ref{sec:phase} we test the robustness of the hub network by performing the same analysis with  data generated with a non-symmetric underlying function at a constant level of noise.  In each case, we generate $300$ data points out of which $50$ are uniformly selected as training points.  The remaining points are used as a test set.  Each network consists of two hidden layers with five neurons per layer and uses sigmoid activation functions.  In order to quantify the robustness of the hub layer in predicting symmetric regression functions, we introduce the even metric: 
\begin{equation}
    \label{eq:evenMetric}
    S_+ = \frac{1}{M}\sum_{i=1}^M\left( \hat x(t_i) - \hat x(-t_i) \right)^2,
\end{equation}
which measures how much a function deviates from the even symmetry,  $S_{+} = 0$ for a perfectly even function.  The index $i$ runs over the set of $M$ test points.   

\subsection{Testing with Data Generated using an Even Function} \label{sec:no-phase}
\begin{figure}[ht!]
    \centering
    \includegraphics[scale=0.32]{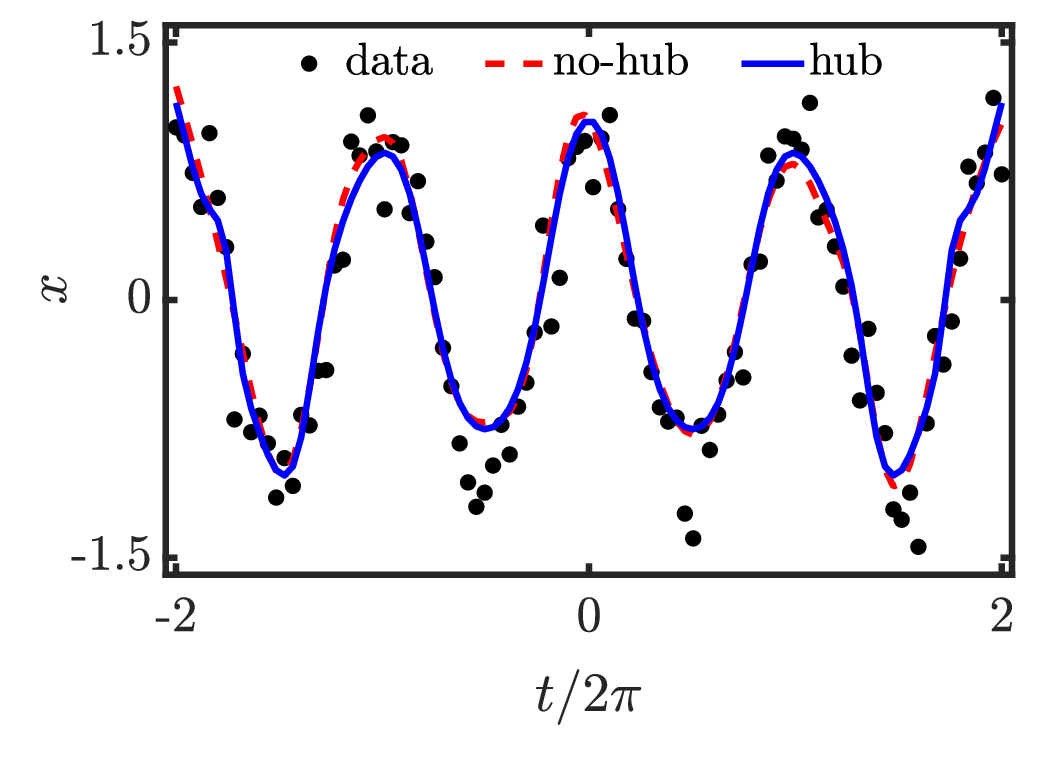}
    \includegraphics[scale=0.32]{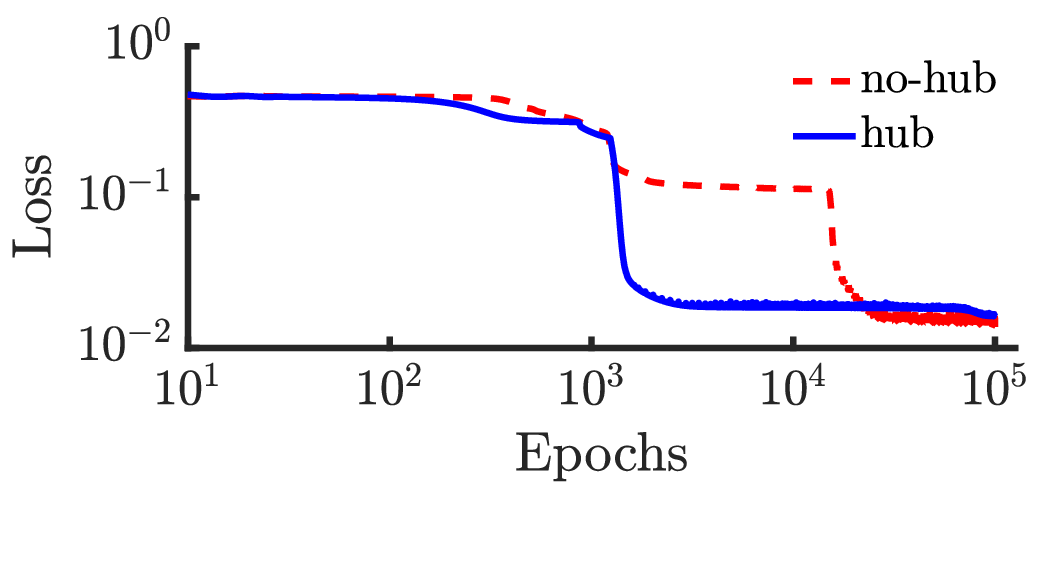}
    \caption{ Left: Regression on noisy data  from an even function. Right:  MSE    in  training.}
    \label{fig:symResults}
\end{figure}

We compare the regression lines predicted from a standard MLP to predictions from the new NN with a hub layer.  Data were generated from Eq.~\eqref{eq:cosSrc}  and $\sigma=0.2$.
In the left panel of Fig. \ref{fig:symResults} the red dashed line is the regression obtained by a standard MLP and the solid, blue line corresponds to an MLP network where the last hidden layer is { replaced} by the even hub layer $H^+$. The right panel of Fig.~\ref{fig:symResults} shows the loss function
on the  training data.  Throughout this work, the loss function is taken to be the mean-squared-error (MSE).  We observe that the loss function converges faster when the hub layer is used.  The MSE of the simple MLP is lower at the end of the training process, which is expected since the simple MLP fits the data where our network compromised the MSE fit in order to preserve the underlying symmetry.

Next, we generate data for different levels of noise for $\sigma\in\left[0,~0.5\right]$ and present the symmetry metric $S_+$ in Fig.~\ref{fig:symResults2} (upper).  
For each $\sigma$, $30$ standard MLP networks were trained and predictions were made with $250$ test points.  Indeed, the even hub layer allows only even functions.  The standard MLP network exhibits a range of $S_{+}$ values, which increases with the noise\footnote{The $S_+$ statistics follows the $\chi^2$ distribution as expected.}.
Moreover, in the same range of $\sigma$ we calculate   the  loss function in the testing data where we observe that the hub layer performs better especially for low level of noise.  The above results are demonstrated by the left panel  in Fig.~\ref{fig:symResults2}.

\subsection{Testing with Data Generated Using Asymmetric Function} \label{sec:phase}
\begin{figure}[ht!]
    \centering
    \includegraphics[scale=0.3]{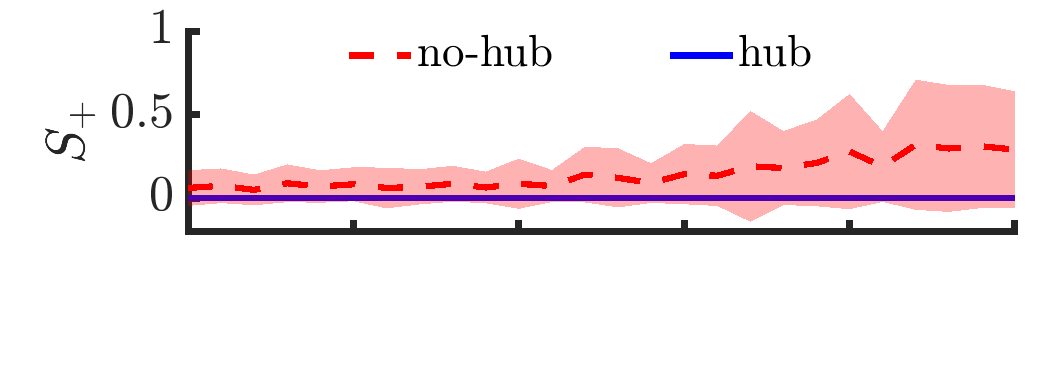} 
    \includegraphics[scale=0.3]{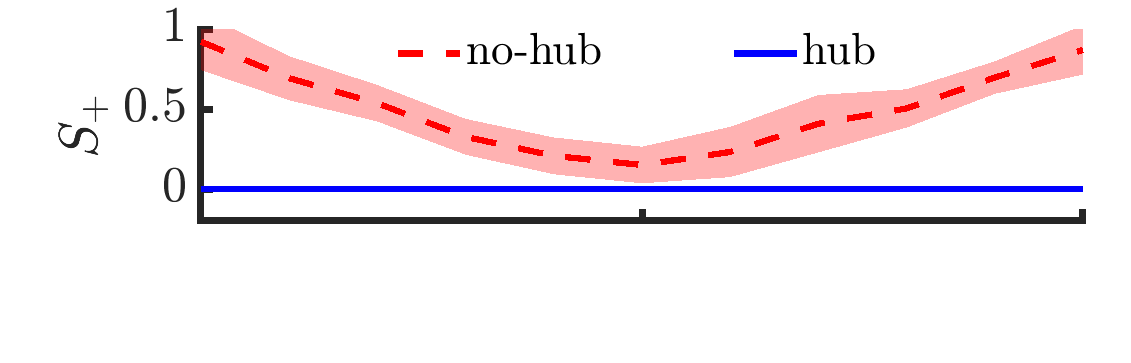}
      \includegraphics[scale=0.3]{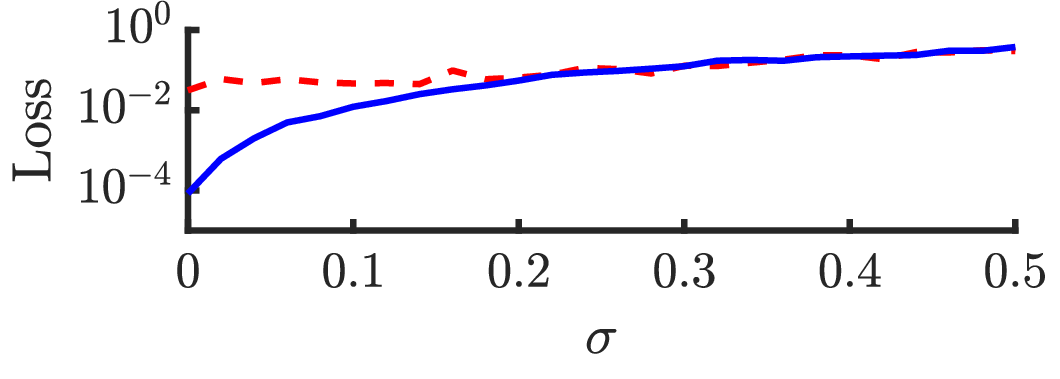}\hspace{.3cm}
    \includegraphics[scale=0.3]{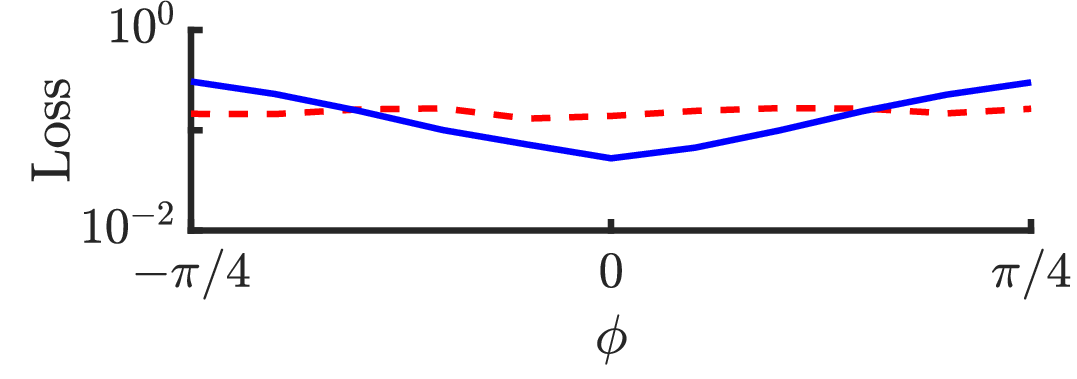}\hspace{.3cm}
    \caption{Even metric and MSE.  Left:   Level of noise. Right: Phase parameter.     }
    \label{fig:symResults2}
\end{figure}

In this case, we generate noisy data with $\sigma=0.2$ and introduce an offset $\phi\in\left[-\pi/4,~\pi/4\right]$ in the even function of Eq.~\eqref{eq:cosSrc} in order to test the robustness of the model. Hence  { the actual data are generated to be asymmetric}. In  Fig.~\ref{fig:symResults2} (right), we again observe that the even hub layer permits only even regression functions ($S_+ = 0$) whereas  for the standard MLP $S_{+} > 0$. As $\phi$ increases  the loss function in the testing set for the hub MLP becomes  larger  than that for the standard MLP because the hub MLP tries to fit an even function but the data are not even symmetric any more, thus  the loss is higher as expected.

In this section, we demonstrated through the toy model of odd/even symmetries that by constraining some of the tunable parameters of the NN we can identically preserve the required symmetries.  This is a general approach, and will be used on a different problems in Section~\ref{sec:energy}.

\section{Constraint Conservation}\label{sec:energy}
Conservation laws are at the heart of many physical systems. {There is a well known theorem in mathematical physics, the Noether's theorem, that establishes the connection between any conservation law and an underlying symmetry of the equations governing the behavior of the physical system \cite{noether}. For example, space translation  symmetry leads to momentum conservation and rotational symmetry yields  angular momentum conservation.}  One of the most celebrated conservation laws is the conservation of energy which derives from the time translational symmetry of the system.  In this section, we show how the idea of the hub neurons introduced in the previous section can leveraged to embed energy conservation in a neural network.  We begin this section with an overview of using neural networks to solve differential equations and a brief review of energy conservation.  We then provide two scenarios in which energy conservation is important.  In subsection~\ref{sec:energy_regression}, a NN is used to predict a function from data that was originally generated from a physical system that conserves energy.  A  hub layer is shown to guarantee, up to the numerical accuracy of the solver, the energy conservation in the subsequent regression.  In subsection~\ref{sec:symplectic}, unsupervised NNs  are used to solve DEs. {Once again, a special symplectic architecture is designed  based on Hamiltonian formulation to embed into a NN  the energy conservation law. Then, the  resulting approximated solutions of the desired system of DEs that obtained by the  symplectic NN conserve the energy.} 

Unsupervised feed-forward NNs offer an alternative approach to the numerical solution of DEs \cite{lagaris1998,spiliopoulos2018,nips2018}.  One key difference between traditional numerical methods and  NNs for solving DEs  is that NNs  seek to learn the actual function that solves the equation, rather than creating an accurate approximation to the function~\cite{spiliopoulos2018}.  A potential advantage of using NNs  to solve DEs  is that the solutions obtained by NNs are differentiable and in a closed, analytic form~\cite{lagaris1998}.  Moreover, NNs for DEs  may be easy to parallelize, thereby leading to significant speed-up in time to solution~\cite{lagaris1998}.

Given a DE  $G\left(t,~ x(t)) \right) = 0$, where $G$ is a (possibly nonlinear) differential operator, the goal is to find $x\left(t\right)$  such that the DE, the initial and the boundary conditions are satisfied.  The unsupervised NN approach proposed in the original work~\cite{lagaris1998} introduces a ``trial'' solution as a re-parameterization of the solution in order to directly satisfy the initial and boundary conditions.  For example, for a first order ODE   with initial condition $x\left(t=0\right) = x_{0}$, the trial solution takes the form $\hat x = x_{0} + t N(t)$ where $N(t)$ is the output from a NN  \cite{lagaris1998}.  In this way, at $t=0$, the trial solution identically satisfies the initial conditions.  The training procedure proceeds as follows \cite{lagaris1998,spiliopoulos2018}:
\begin{enumerate}
  \item Generate  random  points, $t_i$, in the input domain.
  \item Perform a forward pass through the network.
  \item Calculate the loss function: 
    \begin{align}
      L = \sum_{t_i} G\left(t_i, \hat x(t_i) \right)^{2}
    \end{align}
  \item Use back-propagation with stochastic gradient decent to train the network parameters.
\end{enumerate}
Next, we focus on the origin of DEs  and discuss their relationship to conservation of energy.
In mechanics, we often wish to determine how the position of an object evolves in time.  To accomplish this, it is sufficient to know the potential function at  a given position $V\left(\qq\right)$, which characterizes an object's propensity for motion at position $\qq$.  The position $\qq$ is a vector of coordinates (e.g. $ \qq \in {\rm I\!R}^3 $ for three dimension).  We write $q_{k}$ to denote coordinate $k$.  In general, a system consists of many objects, each with their own position vector.  However, to simplify the presentation, we focus on a system consisting of a single object.  The Lagrangian is usually defined as the difference between the kinetic and potential energies,
\begin{align}
  \label{eq:lagrangian}
  \lag(\qq,\dot \qq) = T(\dot \qq) - V(\qq),
\end{align}
where the kinetic energy $T = \dot{\qq}^2 / 2$ quantifies the energy of motion of the object (with mass $m=1$) and dot represents the time derivative of a quantity.  Starting from the Euler-Lagrange (E-L) equation,
\begin{align}
    \label{eq:EL}
    \frac{d}{dt}\left(\frac{\partial \lag}{\partial \dot \qq}\right) - \frac{\partial \lag}{\partial \qq} = 0
\end{align}
the equations of motion are expressed as the differential equations
\begin{align}
  \label{eq:eqn_motion}
 \ddot \qq + \frac{ \partial V(\qq)}{\partial \qq}=0.
\end{align}
Solving these differential equations in Eq.~\eqref{eq:eqn_motion} provides the position of the object as a function of time.  Worth noticing that for a conservative system, the force is negative the derivative of the potential and Eq.~\eqref{eq:eqn_motion} reduces to Newton's law.  Because the focus in this section is on systems that conserve energy, we note that if the Lagrangian does not have an explicit time-dependence, then the energy is conserved.

An alternative formulation of mechanics uses the Hamiltonian, 
\begin{align}
  \label{eq:hamiltonian}
  \ham(\qq, \ppp)  = T(\ppp)  + V(\qq),
\end{align}
where now the kinetic energy $T = \ppp^{2} / 2$ solely depends on the generalized momentum $\ppp$.  The Hamiltonian is related to the Lagrangian via,
\begin{align}
  \label{eq:p_ydot}
  \ppp= \frac{\partial \lag }{ \partial \dot \qq}.
\end{align}
Throughout this work, the potential $V$ is solely a function of $\qq$, which implies 
\begin{align}
  \label{eq:p_eq_qdot}
  \ppp = \dot{\qq}.
\end{align}
Equation~\eqref{eq:p_eq_qdot} will form an essential role in the development of the symplectic NN.  The equations of motion are determined by Hamilton's equations,
\begin{align}
  \dot \qq =~ \frac{\partial \ham}{\partial \ppp}, \hspace{1.09cm}
  \dot \ppp = -\frac{\partial \ham}{\partial \qq}. \label{eq:pdot}
\end{align}
The Hamiltonian itself represents the total energy of the system and in the situation in which the Hamiltonian does not have an explicit dependence on time, the energy is exactly conserved.

In the next section, we assume noisy data that describe the trajectory of a conservative system.  We employ a standard MLP to fit the data and approximate the trajectories which should conserve energy.  The standard MLP architecture fails to accomplish this since it does not know  the underlying conservation law.  We encounter this problem by adding a NN ODE solver that corrects the regression lines to preserve the underlying physics.


\subsection{Energy Conserved in Regression }
\label{sec:energy_regression}
\begin{figure}
    \centering
\includegraphics[scale=0.23]{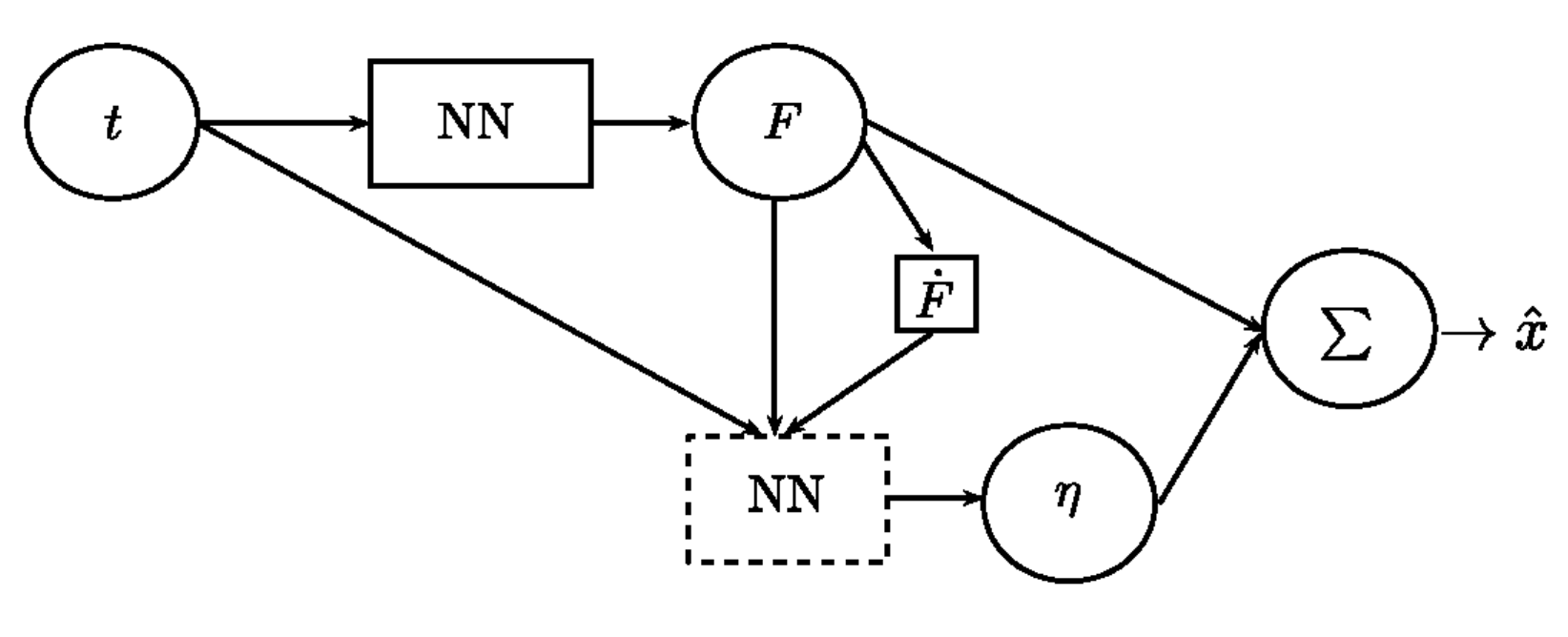}
    \caption{Schematic of a NN  architecture that guarantees a regression that fits data generated from an energy-conserving process.}
    \label{fig:hubODERegNet}
\end{figure}

In this  section, we propose a NN  that  guarantees (limited to the solver accuracy) energy conservation  by perturbing the prediction from an MLP in the correct way.  We work with noisy data that represent the position in time $x(t)$ of an object.  The motion of the object is restricted to one dimension and therefore the coordinates $\qq$ are simply a scalar such that $q = x$.  Moreover, it is assumed that the dynamics describing that motion are energy-conserving.  
We employ a feed-forward MLP with a single linear output node to approximate a regression function $F(t)$ that fits the data by minimizing the MSE.  Since physical laws are not embedded in the MLP structure, $F(t)$ does not represent a function that satisfies energy conservation of the underlying system. We cure this issue by introducing a term $\eta(t)$ that corrects the regression to preserve the energy.  Therefore, the prediction is given by, 
\begin{align}
  \label{eq:x_eta}
  \hat x(t) &= F(t)+\eta(t), 
\end{align}
where $\hat x(t)$ is the prediction from the NN.
From Eq. (\ref{eq:p_eq_qdot}) $\hat p=  \dot {\hat x}$.  We denote total energy of the system by $E$.  Using Eq.~\eqref{eq:hamiltonian} yields,  
\begin{align}
  \label{eq:eta_ode}
  E = \dfrac{1}{2}\dot{\hat{x}}^{2} + V\left(\hat{x}\right)
\end{align}
where $\dot{\hat{x}} = \dot{F} + \dot{\eta}$ from Eq.~\eqref{eq:x_eta}.
The proposed network architecture is depicted in Fig. \ref{fig:hubODERegNet}.  It consists of three parts:
\begin{enumerate}
  \item Use a standard MLP to find $F(t)$  given some input data $t$.
  \item Compute $\dot{F}$ using automatic differentiation and pass $F$ and $\dot F$ to another neural network.  This second NN  solves the first order ODE for the correction $\eta$ implied by Eq.~\eqref{eq:eta_ode},
    \begin{align}
    \label{eq:doteta}
      \dot \eta &= -\dot F \pm \sqrt{2\left(E - V(F+\eta)\right)},
    \end{align}
    The second network is another form of the hub layer introduced in the previous section and used to embed physical constraints in the overall network architecture. { Equation (\ref{eq:doteta}) is a two-value ODE. We turn it in a single value ODE by assuming $\eta\ll F$ and Taylor expanding  Eq. (\ref{eq:doteta}); that yields $\dot \eta\simeq -\dot F$ and thus we replace the $\pm$ with  the  sign of $-\dot F$.}
  \item The outputs from the two NNs  are used to form the final prediction $\hat{x}$.
\end{enumerate}

\subsubsection{Harmonic Oscillator}
As a concrete example, we consider the harmonic oscillator which has the potential,
\begin{equation}
    \label{eq:VforRegr}
    V(x) = \frac{1 }{2} \omega^2 x^2
\end{equation}
where $\omega$ is the natural frequency of the oscillator.  Without loss of generality we take $\omega=1$. 
Solving the E-L equation (\ref{eq:EL}) with  the potential $(\ref{eq:VforRegr})$ and with the initial condition $(x_0, \dot x_0)=(\sqrt{2},0)$  yields the  solution for the position of a harmonic oscillator $x(t)=\sqrt{2}\cos(t)$ and with  energy $E_0=1$.
We generate $100$ data points using the analytical solution in the range $t\in\left[0,4\pi\right]$.  As usual, we introduce Gaussian noise $\epsilon$  with standard deviation $\sigma=0.1$ and mean zero. Hence, the function that generates the noisy data is: $    x(t) = \sqrt{2}\cos(t) + \epsilon$.
We use $50$ of the $100$ data points to train an MLP with a single hidden layer consisting of $20$ sigmoid neurons.
Next, we pass 
$F(t)$ and its derivative to the hub layer, which is a NN ODE solver with one hidden layer consisting of $20$ sigmoid neurons.  The hub layer outputs the correction $\eta(t)$. 

\begin{figure}[ht!]
    \centering
    \includegraphics[scale=0.28]{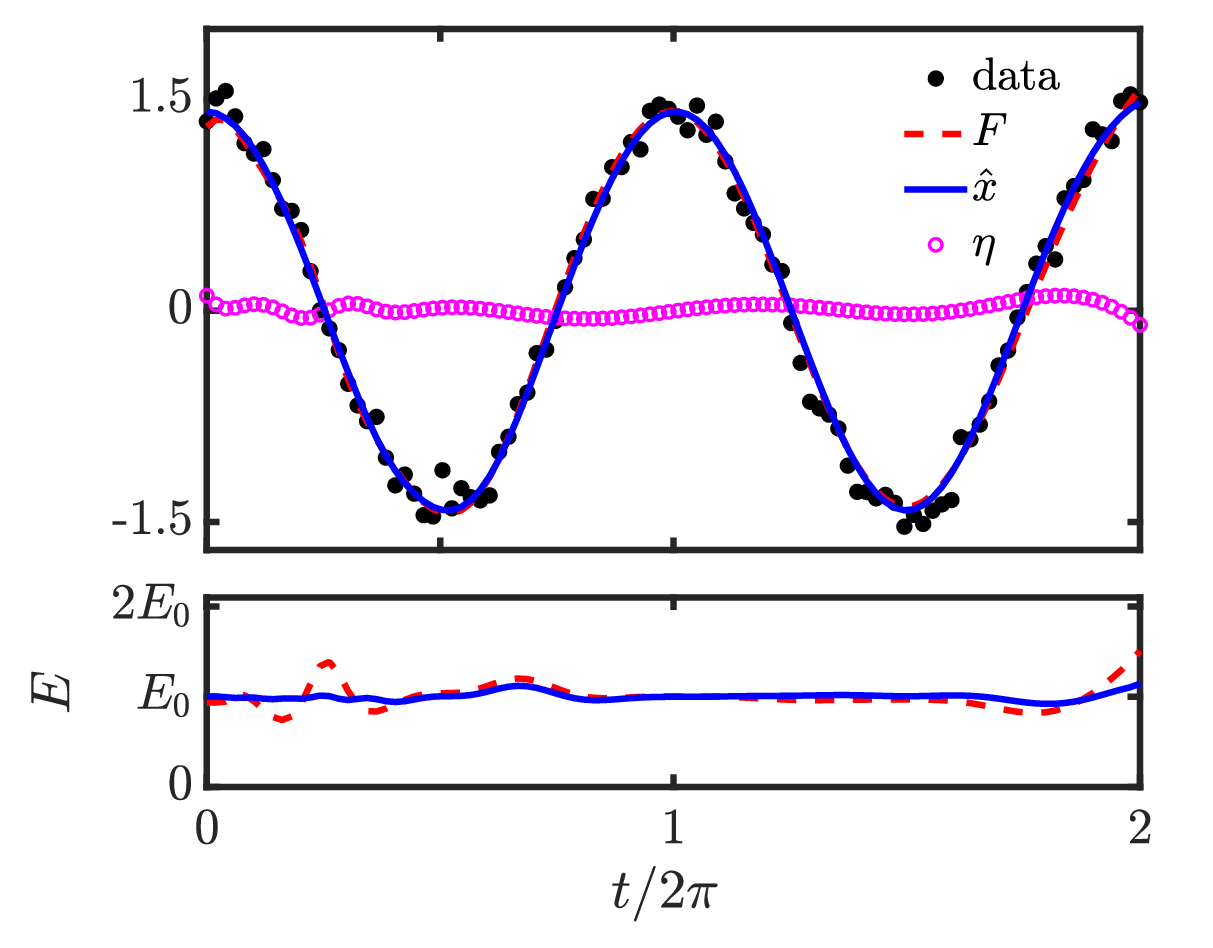}
    \includegraphics[scale=0.28]{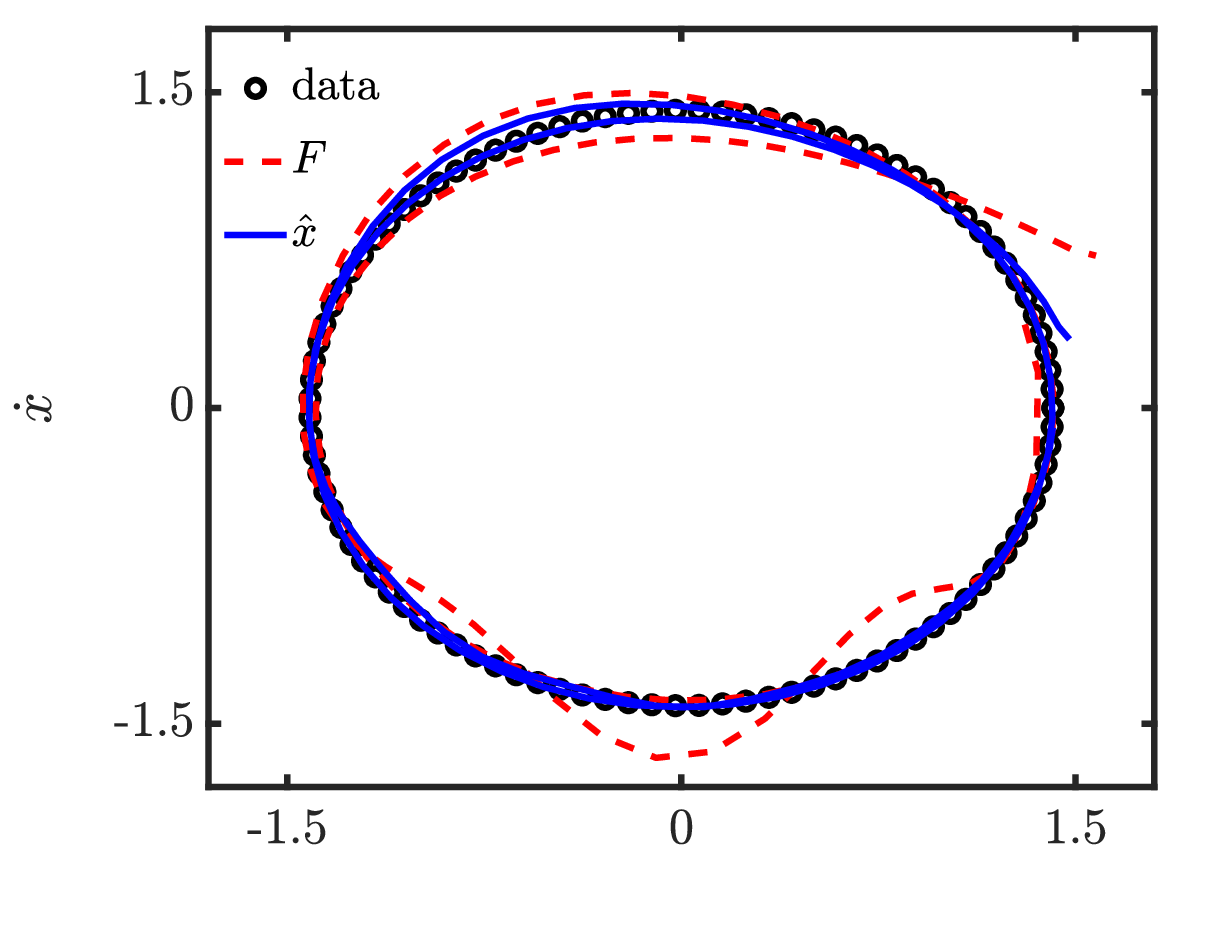}
    \caption {Left: Regression lines and noisy data. Lower: Total energy in time. Right: Phase-space trajectories.     The hub network is able to correct the regression to conserve the total energy and to predict closed trajectories in phase-space.    }
    \label{fig:harmonicOsc}
\end{figure}

Figure \ref{fig:harmonicOsc} presents results from the regression with the standard MLP network and the network with a hub layer for respecting energy conservation.  The left panel shows the actual regression lines predicted from the noisy data.  Both the MLP and the hub network perform well compared to the analytical solution.  We also show the correction $\eta$, which is nonzero.  The lower panel depicts the total energy and shows that the hub network preserves very well the total energy  while the regression from the standard MLP does not. The right panel shows the $\dot{\hat{x}}, \hat{x}$ plot (known as a phase-space plot). We show the exact solution neglecting the noise term which is a closed trajectory.  The hub network  exhibits  closed trajectories (limited again by the NN solver accuracy),   however, the standard MLP fails to capture this essential feature of the system.

\subsection{Symplectic Neural Network}
\label{sec:symplectic}
Differential equations  are used to model the physical world from fluid flows through solid mechanics and materials science. 
These are often  nonlinear equations that are not amenable to analytical techniques. 
 Among the many challenges for numerical methods is the desire that a numerical integrator used for solving the equations somehow respect the intrinsic principles used to derive the equations (e.g. conservation of energy).
 In the present work, we design a symplectic NN architecture for solving  DEs, which guarantees that the predicted solutions preserve the energy. Specifically, we embed constraints in the NN structure that are derived  from Hamilton's equations (Eqs.~\eqref{eq:pdot}).  The constraints reduce the solution space and subsequently, in addition to the correct energy, the NN is more robust and reaches the solution much faster than a non-symplectic architecture. 
 
\begin{figure}[ht!]
    \centering
	 \includegraphics[scale=0.25]{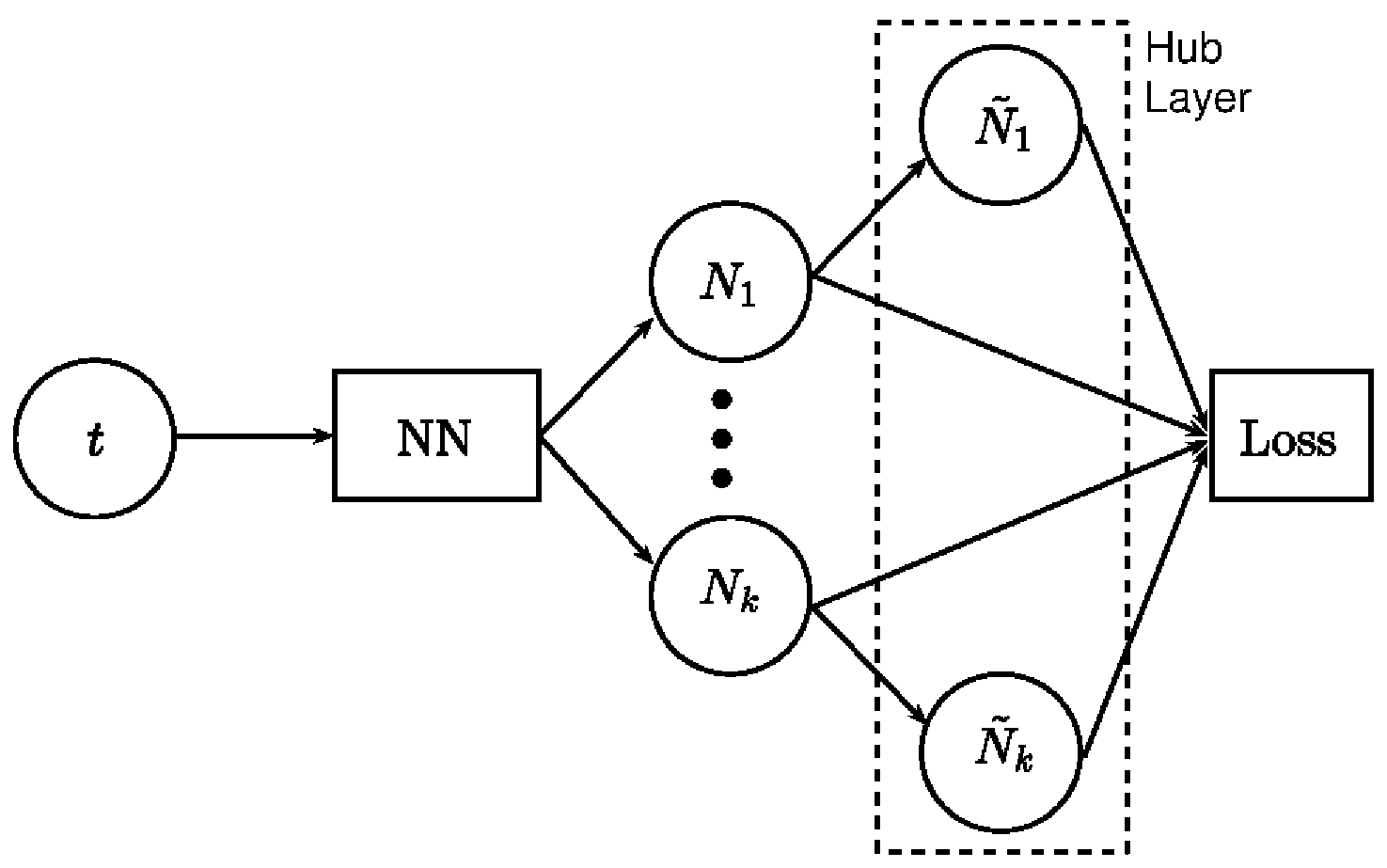}
    \caption{Architecture for solving  ODEs system ensuring the conservation of energy. }
    \label{fig:symplecticStructure}
\end{figure}

We consider a $d$-dimensional system with coordinates $\qq,~\ppp\in {\rm I\!R}^{d}$. Position and momentum coordinate $k$ is denoted by $q_{k}$ and $p_k$, respectively.  The goal is to find the  $\qq$ and $\ppp$ as a function of time as governed by Hamilton's equations (Eqs.~\eqref{eq:pdot}) by using NNs.
Specifically, for each dimension $k$ we have a system of two first order ODEs:
\begin{align}
\label{eq:hamiltonSys}
\dot q_k = p_k, \hspace{1.09cm} \dot p_k = -\frac{\partial V(\qq)}{\partial q_k}.
\end{align}
where we used $\mathcal{H} = {p}_{k}^{2}/ 2 + V\left({\qq}\right)$.
The idea is to design a symplectic NN by imposing the first of the Eqs. (\ref{eq:hamiltonSys}) as a constraint and using the second equation to build the loss function.  For the proposed symplectic NN we suggest the trial solutions which satisfies the initial conditions as
\begin{align}
\label{eq:trial_q1}
\hat q_k &= q_{k,0} + (t-t_0) \ p_{k,0} + \left(1-e^{-(t-t_0)}\right)^2 N_k(t), \\
\label{eq:trial_p1}
\hat p_k &= p_{k,0} +  \left(1-e^{-(t-t_0)}\right) \tilde N_k(t),
\end{align}
where $q_{k,0}$ and $p_{k,0}$ are, respectively, the initial values of position and momentum at $t=t_0$. Imposing the first of the Eqs. of (\ref{eq:hamiltonSys}) we find that 
\begin{align}
\label{eq:tilde_N}
\tilde{N}_k(t) =  \left(1-e^{-(t-t_0)}\right)\dot N_k(t) +2\ e^{-(t-t_0)} N_k(t),
\end{align}
which is a hub neuron. Using Eq. (\ref{eq:tilde_N})  the parametric solution (\ref{eq:trial_p1}) becomes
\begin{align}
\label{eq:trial_p}
\hat p_k &= p_{k,0} +  \left(1-e^{-(t-t_0)}\right)\left[  \left(1-e^{-(t-t_0)}\right)\dot N_k(t) +2\ e^{-(t-t_0)} N_k(t) \right],
\end{align}
which satisfies by structure the initial condition $\hat p_k(t=t_0) = p_{k,0}$ and the first Hamilton'sequation $\dot {\hat q}_k = \hat p_k$. The $N_k(t)$ is the $k^\text{th}$ output of  a feed forward MLP. The time derivative in $\dot N_k(t)$ is obtained by automatic differentiation. The second Hamilton's equation  defines the loss function as
\begin{align}
\label{eq:LossSym}
L =  \sum_k  \left(   \dot{\hat{p}}_{k} + \dfrac{\partial V\left(\hat{\qq}\right)}{\partial \hat{q}_{k}} \right)^2,
\end{align}
which is used for the training of  $N_k(t)$. The proposed symplectic NN is graphically outlined in  Fig.   \ref{fig:symplecticStructure}.

We demonstrate this idea on the H\'{e}non-Heiles (HH) dynamical system \cite{HH_1964}, which was introduced in 1964 to describe the non-linear motion of a star around a galactic center with the motion restricted to a plane.  The HH system has two degrees of freedom such that 
the coordinate variables are $\qq=\left(x,y\right)$ and the momentum variables are $\ppp=(p_{x}, p_{y})$.  
The nonlinear potential of this system is 
\begin{align}
  \label{eq:HHpotential}
  V(x,y) = \frac{1}{2}\left(x^2 +y^2 \right) + \lambda\left( x^2y - \frac{y^3}{3} \right),
\end{align}
where $\lambda$ is a parameter.  The Hamilton's equations results in the nonlinear equations of motion,
\begin{align}
\label{eq:HHq}
      \dot{x} &= p_{x},      &        \dot{y} &= p_{y},    \\ 
\label{eq:HHp}
    \dot{p}_{x} &= -\left( x + 2\lambda x y \right), &  \dot{p}_{y} &= -\left( y + \lambda \left(x^{2} - y^{2} \right) \right).
\end{align}
The Eqs. (\ref{eq:HHq}) are embedded in our proposed NN.  The Eqs. (\ref{eq:HHp}) define the loss function according to  Eq. (\ref{eq:LossSym}) as 
\begin{align}
\label{eq:LossHH}
L =  \left( \dot{\hat p}_{x} +  \hat x + 2\lambda \hat x \hat y  \right)^2  +  \left( \dot{\hat p}_{y} +  \hat y + \lambda \left(\hat x^{2} - \hat y^{2} \right)  \right)^2,
\end{align}
which is used for training an MLP NN with two outputs $(N_x,\ N_y)$. The quantities $\hat x, \hat y$ and $\hat p_x, \hat p_y$ are determined by the trial solutions (\ref{eq:trial_q1}) and (\ref{eq:trial_p}), respectively.

We apply the symplectic NN architecture  to find the solution of the HH dynamical system in a chaotic regime.  The initial conditions for the simulation are $\left(x_0,y_0,p_{x,0},p_{y,0}\right) = (0.3, -0.3, 0.3, 0.15)$ and for $\lambda = 1$,  corresponding to the  energy $E_{0} = 0.13$.  We reiterate that due to the symplectic structure of the NN, the energy should remain constant for all time.  The NN  is trained on a random grid of $150$ points for $t\in\left(0,10\pi\right)$, which is randomly selected at the beginning of each epoch \cite{spiliopoulos2018}. We use two hidden layers with $N=50$ nodes   in each hidden layer. For the nonlinear activation function of each hidden node we choose the trigonometric function $\sin()$. 
We compare the solutions obtained from the symplectic NN to those obtained by a standard MLP using the same network hyper-parameters and using sigmoid activation functions for the hidden nodes.   For the standard MLP, instead of solving for the dynamics from Hamilton's equations, we solve the system of second order equations for $x$ and $y$ following the E-L Eq. (\ref{eq:eqn_motion}):
\begin{align}
  \ddot x +x +2\lambda x y =0, \hspace{2cm} 
  \ddot y + y +\lambda(x^2-y^2) =0,
\end{align}
and with the trial solutions \cite{lagaris1998}
\begin{align}
  \hat x &= x_0 + (t-t_0)p_{x,0} + (t-t_0)^2 N_x^\text{MPL},\\
  \hat y &= y_0 + (t-t_0)p_{y,0} + (t-t_0)^2 N_{y}^\text{MPL},
\end{align}
where $N_x^\text{MLP}$ and $N_y^\text{MLP}$ are the two outputs from a standard MLP NN. This corresponds to the standard, physics-agnostic approach.
Finally, we also compare the NN solutions to solutions obtained using an adaptive, non-symplectic time-integrator (\texttt{odeint} in the Scipy package of python).

\begin{figure}
    \centering
\includegraphics[scale=0.3]{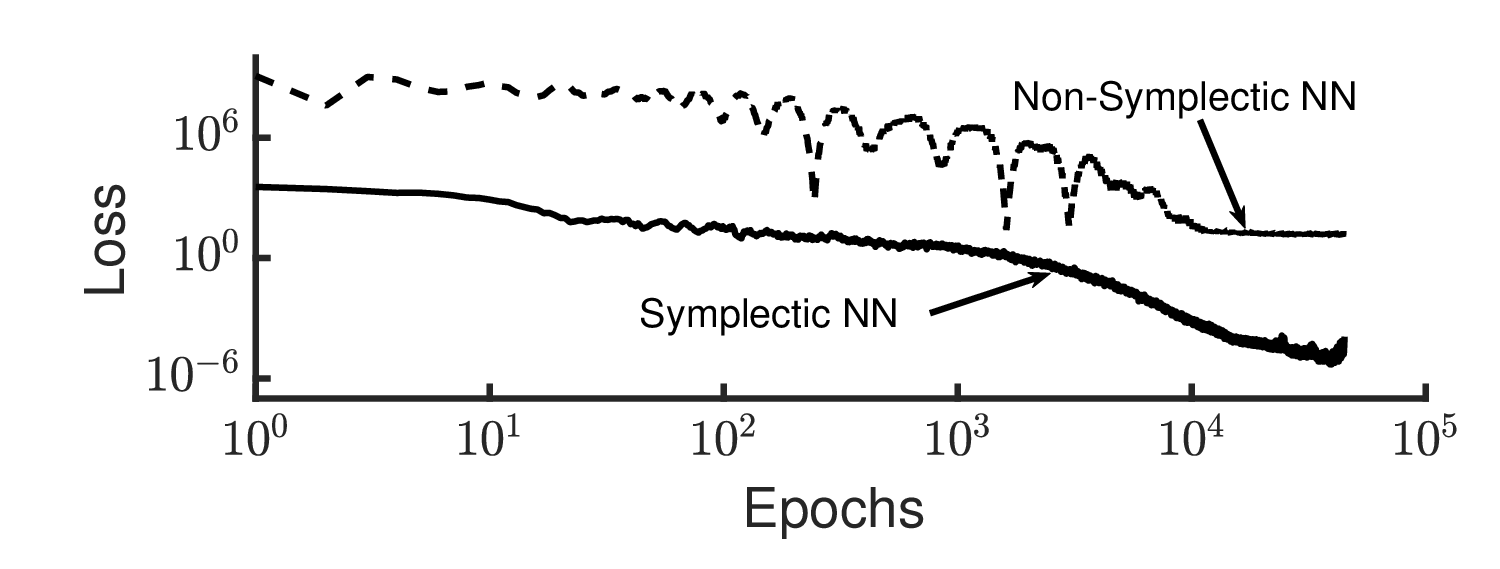}
    \caption{Training loss of the  symplectic and non-symplectic NN in solving the HH system.}
    \label{fig:odeLoss}
\end{figure}

The loss function for the symplectic and standard MLP  NNs during the training process is represented by  Fig.~\ref{fig:odeLoss}. The symplectic NN loss is lower, more uniform, and faster in convergence than the loss function for the standard MLP.

\begin{figure}[ht!]
    \centering
    \includegraphics[scale=0.29]{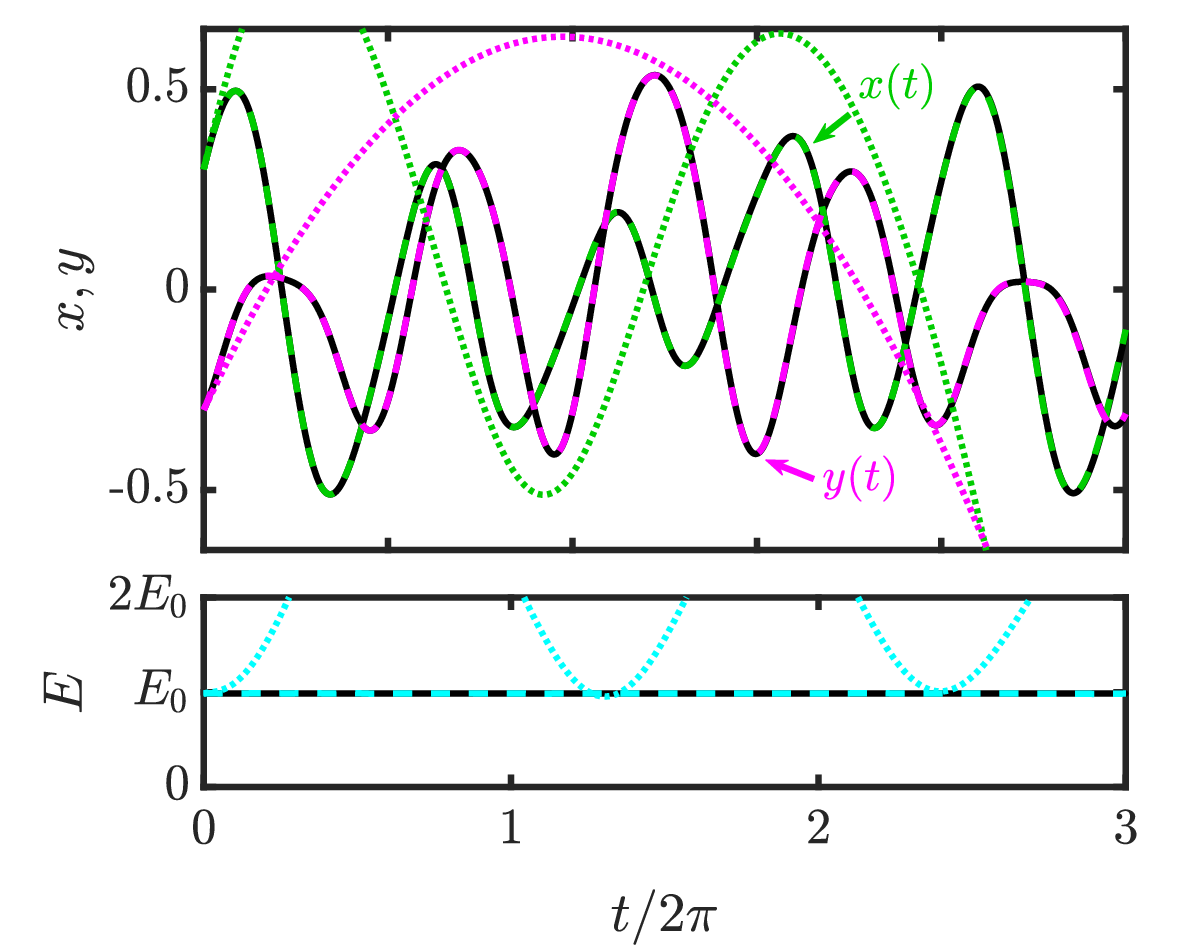}
    \includegraphics[scale=0.29]{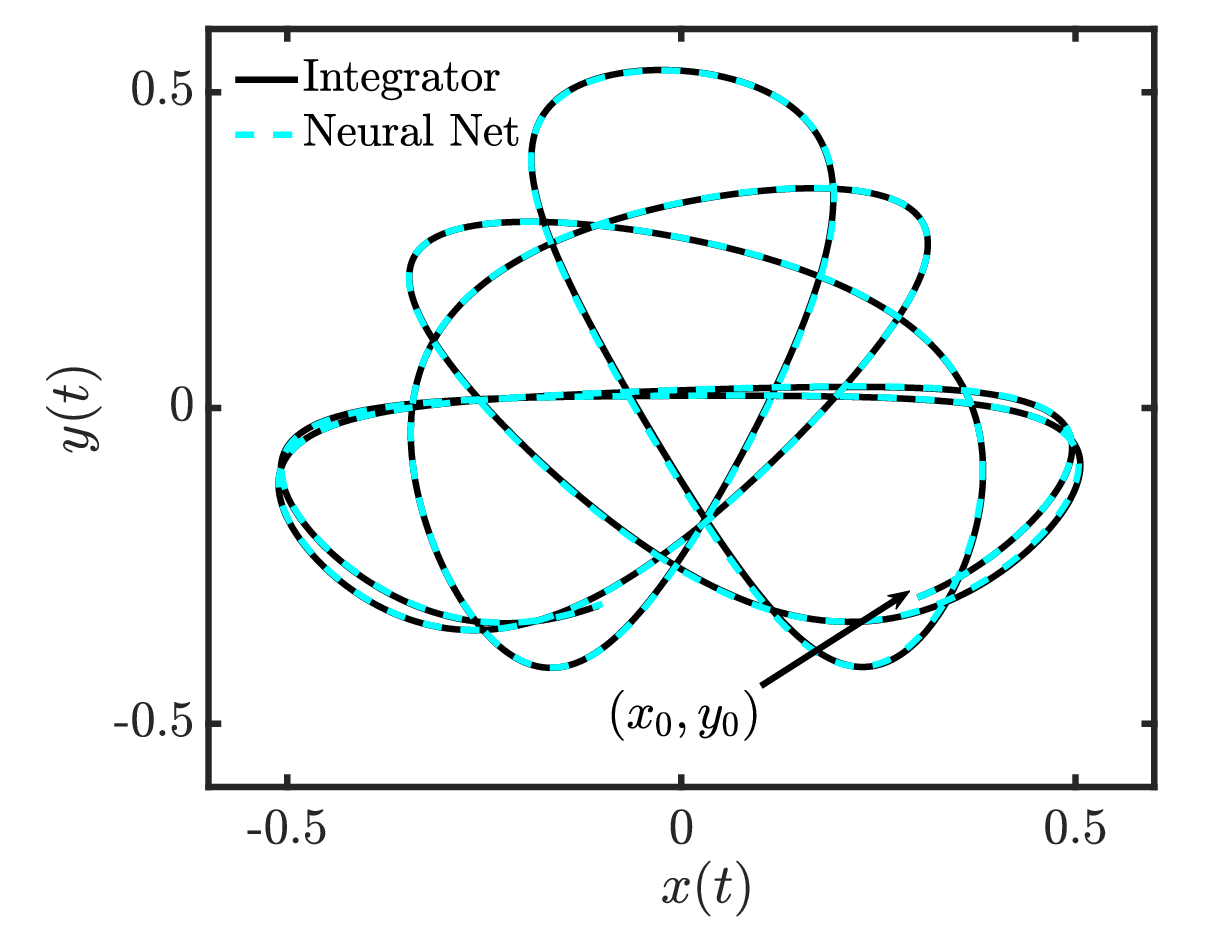}
    \caption{Left: Solutions  by a numerical integrator (dashed),  symplectic NN (solid), and non-symplectic NN (dotted). Lower: Energy in time. Right: Orbit in $x-y$ plane.}
    \label{fig:HH}
\end{figure}

Figure~\ref{fig:HH} (left) shows the solution for the position $\qq$ as a function of time.  The solid black line represents the numerical solution and the dashed color lines represent the solution from the symplectic NN. 
The standard MLP solution, which uses the same hyperparameters as the symplectic network, is shown in the color dotted lines. The symplectic network outperforms the standard MLP, which completely fails to predict the correct solution behavior.  We note that the standard MLP solution can be improved by adjusting the hyperparameters (e.g. number of neurons) or by training the network for orders of magnitude more epochs than that used to train the symplectic network.
The dashed  blue line in the lower panel of Fig.~\ref{fig:HH} shows that the energy remains constant in time thereby demonstrating the energy conservation property of the symplectic network. {The dotted line shows that the standard MLP does not conserve energy.}  Finally, the right panel of Fig.~\ref{fig:HH} shows the predicted chaotic orbit in the $x-y$ plane. 

\section{Conclusion}
\label{sec:conclusion}
In recent years, machine learning has made in-roads in traditional science and engineering fields.  Since these methods are relatively new to physics, there are many outstanding issues.  In this paper, we proposed a neural network paradigm, called the hub neurons (or hub layer), that provides the capability to embed physical constraints into neural networks.  This approach goes beyond the standard approach of introducing regularization terms to enforce such constraints.  Moreover, it provides a mechanism to exactly satisfy known physics such as conservation laws.  Such a property is crucial if neural networks are to be used for predictive science, especially if they are employed in regimes for which data is unavailable.  In addition to predictive power, networks with embedded physics are faster to train and may be more robust to sparse and noisy data sets.  We tested the hub layer concept on three different physical constraints.  First, we designed a hub layer neural network to account for even/odd symmetry.  The new network outperforms a standard neural network while simultaneously exactly preserving the desired symmetry.  Next, we focused on conservation of energy using noisy data that was generated from an energy-conserving process.  By introducing a hub layer into the neural network that corrects the regression based on the underlying Hamiltonian dynamics, the predictions from the new neural network show a great improvement over a classical neural network.  Finally, we use an unsupervised neural network to solve a system of nonlinear differential equations, which conserves energy.  Again, we use the hub neuron concept to design a symplectic neural network that guarantees that the neural network predictions are consistent with energy conservation.  Giving these promising results, we plan to embed other important symmetries into neural networks in the future.

\subsubsection*{Acknowledgments:} 
MM  acknowledges  useful discussions with Prof. G. P. Tsironis. MM and EK acknowledge  partial support from EFRI 2-DARE NSF Grant No. 1542807 and  from ARO MURI Award No. W911NF14-0247. We used computational resources on the Odyssey cluster of the FAS Research Computing Group at Harvard University.

\end{document}